\documentstyle[12pt]{article}

\def\onehalfspace{\baselineskip=18pt}

\begin{document}

\onehalfspace

\begin{center}
{\large {\bf The Continuum Slopes of Optically Selected QSOs}}
\end{center}

\begin{center}
Paul J. Francis \\ 
School of Physics, University of Melbourne, Parkville \\ 
Victoria 3052, Australia.\\ E-mail: pfrancis@physics.unimelb.edu.au 
\end{center} 
\medskip

\medskip

\noindent
{\it Keywords:} galaxies: quasars: general

\medskip 

\noindent
{\bf Abstract:} Quasi-simultaneous optical/near-IR photometry is presented 
for a sample of 37 luminous optically selected QSOs drawn from the Large
Bright QSO Survey. Most of the QSOs have decreased in brightness
since discovery; this is expected in flux-limited samples. The continuum
shape of most of the QSOs can be represented by a power-law of the
form $F(\nu) \propto \nu^{-0.3}$, but a few have softer (redder) continuum 
slopes.

\medskip

\noindent {\bf 1. Introduction}

The shape of the continuum spectra of AGN in the rest-frame UV-optical region
has long been uncertain. It is typically parameterised as a power-law
of index $\alpha$, where the flux per unit frequency $f_{\nu} \propto
\nu^{\alpha}$. The mean value of $\alpha$, the dispersion in values of
$\alpha$, and even the validity of this power-law parameterisation, are all
controversial.

For many years, a mean value of $\alpha = -0.7$ has been assumed by many
researchers, based largely on the observations of Richstone \&
Schmidt (1980) (see also O'Brien, Gondhalekar \& Wilson 1988 and Sargent,
Steidel \& Boksenberg 1989).
Recently, however, a number of authors have claimed that the true mean
continuum slope is actually much harder (bluer) than this, with $\alpha = 
-0.3$ (eg. Neugebauer et al. 1987, Francis et al. 1991). There is equal
disagreement about the range of continuum slopes shown by QSOs, with
some authors claiming that all QSOs have essentially the same continuum
slope (ie. little or no dispersion in $\alpha$, Sanders et al. 1989) while
others claim that the dispersion in continuum slopes is large
(Elvis et al. 1994, Webster et al. 1995).
 
These disagreements are important both physically, and because of the
enormous uncertainty they introduce into measurements of QSO evolution.
If, for example, the mean value of $\alpha$ is $-0.3$, rather than $-0.5$
as commonly assumed, the number counts of bright QSOs at $z \sim 2$ can be
explained with only half the evolution in the comoving density of luminous 
QSOs previously required, due to the
difference in $k$-corrections. Allowing for a reasonable scatter in
continuum slopes can also introduce factors of two or more into the
inferred evolution of QSO number densities (Giallongo \& Vagnetti 1992, 
Francis 1993b).

Physically, a value of $\alpha \sim -0.3$ is consistent with the
free-free emission models of Barvainis (1993), whereas a large dispersion
in continuum slopes, unless caused by dust, would rule out this model.
Webster et al. (1995) claim that flat-spectrum radio-loud quasars show
a much greater dispersion in their continuum slopes, and a much softer
(redder) mean slope than optically selected QSOs, and ascribe this difference
to dust obscuration, a claim disputed by Serjeant \& Rawlings 1996, who ascribe
the results to relativistically beamed red synchrotron emission.

The different values of $\alpha$ reported in the literature may reflect 
real differences between different classes of QSOs, such as radio-loud 
and radio-quiet, or luminous and faint. It is however possible that the
differences reflect no more than different measurement techniques. One
problem is that most AGNs with multi-wavelength observations are low 
luminosity, so the contribution of the host galaxy light is
important (Elvis et al. 1994) and needs to be carefully corrected for.
Measurement in the optical is complicated by the many strong blended
Fe~II emission-lines that obscure large parts of the spectra (Wills,
Netzer \& Wills 1985, Francis et al. 1991). This obscuration makes
measurements of continuum slopes inferred from optical data alone very 
unreliable, a problem that can be solved by the use of multi-waveband
data. Alas, such data is seldom taken simultaneously, allowing
variability to affect the inferred continuum slopes.

In this paper, an attempt is made to provide a definitive measurement of
the continuum slope distribution of one type of quasars; luminous
optically selected QSOs. The QSOs are drawn from the Large Bright
QSO Survey (LBQS, Hewett, Foltz \& Chaffee 1995), a large, uniform
and highly complete optically selected QSO sample that is increasingly
being used as a standard against which to compare other quasar samples
(eg. Webster et al. 1995). These QSOs are bright enough that the emission 
from their host galaxies can be neglected, as shown below. To minimise 
the effect of blended emission-lines, the continuum slopes are inferred 
from multi-waveband optical and near-IR photometry covering a factor
of five in wavelength. The photometry is 
quasi-simultaneous, to minimise any variability effects. This data set 
will provide a benchmark against which the continuum
slopes of other samples can be compared.

The sample and observations are described in Section 2, the results 
are presented and discussed in Section 3, and conclusions drawn in Section 4.

\medskip

\noindent {\bf 2. Observations}

\medskip

\noindent {\it 2a. The Sample}

\medskip

A subset of QSOs drawn from the Large Bright QSO Survey (LBQS) were
observed. The LBQS is described in Foltz et al. (1987, 1989), Hewett et al.
(1991), Chaffee et al. (1991) and Morris et al. (1991), and summarised in
Hewett, Foltz and Chaffee (1995). It consists of 1065 QSOs, with redshifts
$0.2 < z < 3.4$, apparent magnitudes $16.0 < B_J < 18.85$, and absolute
magnitudes $M_{B_J} < -21.5$ (for $q_0 = 0.5$, $H_0 = 100 
{\rm km\ s}^{-1}{\rm Mpc}^{-1}$). The QSOs were selected from UK Schmidt
objective prism plates, using a variety of automated selection criteria,
including continuum colour and the presence of strong emission-lines
and continuum breaks. Comparison with existing surveys suggests that the
LBQS is highly complete to its $B_J$ magnitude limit. Radio observations
of a sub-set of the LBQS are presented by Visnovsky et al. (1992), Francis,
Hooper \& Impey (1993) and Hooper et al. (1995). Discovery epoch $B_J$
magnitudes were obtained by scans of UK Schmidt plates and are accurate
to $\sim 0.15$ mag.

$B$, $V$, $H$ and $K$ photometry for subset of 37 LBQS QSOs is presented
in this paper. The subset observed were selected to be optically
bright, high redshift, and accessible during the observing runs.

At redshifts $z \sim 2$, typical of most of the subsample observed, the 
$K$-band corresponds to the rest-frame $R$ band, and the $B$-band corresponds
to rest-frame 1500\AA . We are therefore measuring the slope of the 
long-wavelength end of the `Big Blue Bump' (Sanders et al. 1989).

The spectral synthesis models of Bruzual
\& Charlot (1993), and of Rocca-Volmerange \& Guiderdoni (1988), were used
to estimate the colours of the most luminous plausible host galaxies;
starbursting giant proto-ellipticals. The rest-frame  rest-frame $R$-band 
magnitudes at $z \sim 2$ were be calculated, and converted into observed 
$K$-band magnitudes. Even the most extreme galaxies had $K >17.5$; two
magnitudes fainter than the QSOs in this sub-sample (Table~1). Host galaxy 
light is thus unlikely to contribute significantly to these observations.
 
\medskip

\noindent {\it 2b. Observations}

\medskip

Optical photometry, in the $B$ and $V$ bands, was obtained on the
Steward Observatory 90'' telescope on the night of 3rd September 1992.
Conditions were photometric, but due to malfunctions in both the
autoguider and the tracking, fast manual guiding was required. Five minute
CCD exposures were obtained. Not all objects could be imaged, due to
substantial overheads associated with the manual guiding.

Near-IR photometry, in the $H$ and $K_s$ bands, was obtained on the
nights of 27th and 28th April 1992, and 27th -- 30th September 1992, on the
Steward Observatory 61'' telescope, using the 256 $\times$ 256 NICMOS 
HgCdTe array camera. 3 $\times$ 3 raster patterns were
used, with three 50 second exposures at each grid pointing in $H$, and
five 30 second exposures in $K_s$ to give an exposure time of 22.5 minutes.
Conditions were photometric throughout.

The data was reduced using standard IRAF\footnote{IRAF is distributed
by the National Optical Astronomy Observatories, which is operated by
the Association of Universities for Research in Astronomy, Inc. (AURA)
under cooperative agreement with the National Science Foundation.} procedures.
Magnitudes were measured using very large fixed apertures, as sky noise
was not the dominant uncertainty, and the point spread function was very
irregular due to the manual autoguiding.

Only the September data was used to derive continuum slopes; the optical
and IR photometry are separated by three weeks, and so are not strictly
simultaneous. Optically selected QSOs, however, show little variability of
timescales of less than a month (Cristiani et al. 1996). Magnitude errors
in both the optical and near-IR are $\sim 0.05$ magnitudes.

\medskip

\noindent {\bf 3. Results and Discussion}

\medskip

The results of the photometry are shown in Table~1.

\medskip

\noindent {\it 3a. Variability}

\medskip

The UK Schmidt sky survey plates from which the original magnitudes
were measured were taken in the late 1970s and early 1980s. The
time interval between the original $B_J$ magnitudes and our $B$ and $V$
photometry is thus $\sim 10$ years, allowing the assessment of the optical
variability of the sample.

1992 epoch $B_J$ magnitudes were calculated using the
approximate relationship $B_J = 0.7 B + 0.3 V$, and the change in
$B_J$ between the Schmidt plate epoch and 1992 calculated. A histogram
of the changes is shown in Figure~1.

As Figure 1 shows, the magnitude changes are very lopsided; of the 26
QSOs measured, 22 decreased in brightness over the $\sim 10$ years, while
only 4 increased in brightness. The situation is even more lopsided when
larger magnitude changes are considered; of the 16 QSOs which varied
by more that $0.1$ magnitudes, none increased in brightness. The median
change in brightness is a decrease of $0.17$ magnitudes, and the standard
deviation is $0.23$ magnitudes. This standard deviation 
is in good agreement with other studies (eg. Cristiani et al. 1996, and
references therein). This lopsidedness remains regardless of the exact
photometric techniques employed.

Why are the magnitude changes so lopsided? If QSO variation is a random walk
process, the probability of a QSO increasing and decreasing in brightness
will be equal. If this were true, the probability of getting such a lopsided
distribution by chance is $0.6$\%.

If, however, the variability of a QSO is not a random walk process, but
each QSO has some characteristic brightness around which it varies, the
lopsidedness can be explained. Any sample will consist of QSOs that were
abnormally bright at the time of selection, and those that were abnormally
faint. The QSO luminosity function is, however, very steep, so a QSO
detected at a given magnitude is far more likely to be a fainter QSO
caught in a bright state, than a brighter QSO caught in a faint state.

The size of the expected lopsidedness can be crudely estimated as follows.
The typical variation is around $0.2$ mag. Given the luminosity function of
bright QSOs (eg. Boyle, Shanks and Peterson 1988), QSOs $0.2$ magnitudes
brighter than a given magnitude are $\sim \times 4$ less common than those
$0.2$ magnitudes fainter than the given magnitude. Thus the number
of QSOs at the given magnitude which were at a maximum at the selection
epoch is likely to exceed that which were at a minimum by a factor
of $\sim 4$. Thus the ratio of the number QSOs that subsequently grow fainter 
to the number that grow brighter should be $\sim $ 4:1, in rough agreement 
to that observed.

\medskip

\noindent {\it 3b. Continuum Slopes}
 
\medskip

Continuum slopes were computed as follows. Only QSOs with both optical
and near-IR photometry were used. The optical magnitudes were corrected for
the presence of strong broad emission-lines in the $B$ and $V$ bandpasses.
For each spectrum, the redshift determined which strong emission-lines
would be in each filter bandpass. If a strong line was present, its equivalent
width in that QSO was obtained from the measurements of Francis (1993a). 
The flux in the filter bandpass was then corrected
for the presence of this line, allowing for the redshift dependence
of observed-frame equivalent widths. For the near-IR bandpasses, which 
often contained H-$\alpha$ and H-$\beta$, no measurements of the equivalent
widths of individual QSOs were available, so the fluxes were corrected by 
assuming a mean rest-frame equivalent width of 58\AA\ for H-$\beta$ and 
290\AA\ for H-$\alpha$ (Francis et al. 1991).

For the higher redshift QSOs, Ly-$\alpha$ forest absorption can reduce
the $B$-band flux. At $z>3$, the whole $B$-band is affected by
Ly-$\alpha$ absorption, and no correction was attempted; the $B$-band
flux was simply not used in the fit. For $2.4 < z < 3$ however, the
$B$-band flux was corrected, the correction factor being chosen from 
visual inspection of the spectrum. Magnitude corrections (corrected
magnitude minus observed magnitude) are shown in Table~2, and the
corrected continuum fluxes in Table~3. Note that these corrections are
typically small; even if no corrections are employed, the results of
this paper are almost identical.

No attempt has been made to correct for weak emission-lines, and for
Fe~II emission. The use of a mean equivalent width for the Balmer
line correction also introduces an error of $\sim 30$\% in the
$H$ and $K$-band correction factors (Francis 1993a). However, the 
long wavelength
baseline ($0.4$ -- $2.2 \mu$m) means that these errors will not have
a large effect on the derived continuum slope $\alpha$; a 25\% error
in one of the continuum fluxes introduces only an error of $\sim 0.1$
in the continuum slope index $\alpha$. Errors in $\alpha$ are
predominantly systematic and hard to estimate, but from the
self consistency of objects for which we have multiple observations,
and the residuals to power-law fits to the continuum (Table~3), we estimate
them to be of order $0.1$.

A power-law continuum is fitted (by least-squares minimisation) to the 
corrected continuum fluxes; the resultant power-law indicies $\alpha$
are listed in Table~3. A histogram of these continuum slopes in shown in
Figure~2. The power-law fit is generally a good one; the scatter around
the fits (Table~3) is consistent with the photometry and correction errors. 
Note however
that in the absence of $R$ or $I$-band photometry, we have little sensitivity
to large-scale departures from a power-law continuum shape.

The mean value of the continuum slope index $\alpha$ is $-0.46$, with
a root-mean-squared (rms) dispersion of $0.30$. Thus the mean slope is
harder than the canonical $\alpha = -0.7$, but not as hard as the
$\alpha = -0.3$ claimed by Francis (1993b). Note however that the
median slope is $-0.35$ and the mode $\sim 0.3$; the mean being
biassed towards higher values by the two outlying points. There are
no obvious observational problems with the observations of these two
outlying points; their redness is much too large to be explained by errors
in the photometry and/or correction factors. Interestingly, both the 
outlying points are radio-loud QSOs, though the sample is too small to 
draw any significant conclusions.

These results demonstrate that the continuum slopes do show a real dispersion,
but a much smaller one than that claimed by Webster et al. (1995) and
Elvis et al. 1994). A standard deviation of $0.3$ in $\alpha$ agrees with
that found by Giallongo \& Vagnetti (1992), who showed that it  leads to
important consequences for measurements of QSO luminosity functions. The
hard (blue) mean continuum slope reported here also suggests that previous
workers, using $\alpha = -0.7$, will have overestimated the evolution
of the number density of QSOs. A dispersion in $\alpha$ of $0.3$ is smaller
than that reported by Francis (1993b) for the LBQS as a whole ($0.5$);
the difference is probably that the measurements here are more accurate, 
being based on a much wider wavelength coverage.

\medskip

\noindent {\bf 4. Conclusions}

\medskip

The very hard (blue) modal continuum slope of this sample agrees well
with the measurements of Sanders et al. (1989), for another sample
of luminous, optically selected QSOs. It is however, very different from
the continuum slopes of the radio-selected sample of Webster et al. (1995)
and the heterogeneous sample of Elvis et al. (1994), both of which show
a substantial dispersion in continuum slopes, with $-0.3 > \alpha > -4$.
It is interesting to note, however, that the continuum slopes of the
optically selected QSOs lie at the blue envelope of the continuum
slope distribution of Webster et al.'s radio selected quasars.

Webster et al. suggested that the redness of their data was caused by
dust obscuration, somewhere along the line of sight, and  that the true
continuum shape of QSOs is a uniform $f(\nu) \propto \nu^{-0.3}$. They
showed that if this were true for radio-quiet QSOs, optically magnitude 
limited samples would show a distribution of continuum slopes with a narrow
peak at $\alpha = -0.3$, and a low tail to redder slopes. This is a
good description of our data (though given the small sample size,
many equivalently good descriptions are possible).

Our observations are therefore {\em consistent} with a picture in which 
quasars have an intrinsic continuum slope of $\alpha \sim -0.3$ (as
predicted by the free-free emission model), but
are reddened by various amounts of dust, as suggested by Webster et al.
Most of the heavily reddened quasars will be missed from samples, 
such as the LBQS, with a blue magnitude limit, but a few moderately red
sources will remain, as observed. This model will be further elaborated in
Francis et al. (1996, in preparation).

\medskip

\noindent {\bf Acknowlegements}

\medskip

I wish to thank George and Marcia Rieke for their assistance in preparing for
and reducing the IR observations, and Rachel Perkins for her help at 
the telescope. I acknowledge support from a NATO advanced fellowship, an 
ARC large grant, and National Science Foundation grant AST 90-01181.

\newpage

\noindent{\bf References}

\medskip
\noindent
Barvainis, R. 1993, ApJ, 412, 513

\medskip
\noindent
Boyle, B. J., Shanks, T., \& Peterson, B. A. 1988, MNRAS, 235, 935

\medskip
\noindent
Bruzual, A. G., \& Charlot, S. 1993, ApJ, 405, 538

\medskip
\noindent
Chaffee, F. H., Foltz, C. B., Hewett, P. C., Francis, P. J., Weymann, R. J.,
Morris, S. L., Anderson, S. F., \& MacAlpine, G. M. 1991, AJ, 102, 461

\medskip
\noindent
Cristiani, S., Trentini, S., La Franca, F., Aretxaga, I., Andreani, P., Vio, R.,
\& Gemmo, A. 1996, A \& A, 30, 395

\medskip
\noindent
Elvis, M., Wilkes, B. J., McDowell, J. C., Green, R. F., Bechtold, J.,
Willner, S. P., Oey, M. S., Polomski, E., \& Cutri, R. 1994, ApJS, 95, 1

\medskip
\noindent
Foltz, C. B., Chaffee, F. H., Hewett, P. C., MacAlpine, G. M., Turnshek, D. A.,
Weymann, R. J., \& Anderson, S. F. 1987, AJ, 94, 1423

\medskip
\noindent
Foltz, C. B., Chaffee, F. H., Hewett, P. C., Weymann, R. J., Anderson, S. F.,
\& MacAlpine, G. M. 1989, AJ, 98, 1959

\medskip
\noindent
Francis, P. J. 1993a, ApJ, 405, 119

\medskip
\noindent
Francis, P. J. 1993b, ApJ, 407, 519

\medskip 
\noindent 
Francis, P. J., Hewett, P. C., Foltz, C. B., Chaffee, F. H., Weymann, R. J. 
\& Morris, S. L., 1991, ApJ, 373, 465 

\medskip
\noindent
Francis, P. J., Hooper, E. J. \& Impey, C. D. 1993, AJ, 106, 417

\medskip
\noindent
Giallongo, E., \& Vagnetti, F. 1992, ApJ, 396, 411

\medskip
\noindent
Hewett, P. C., Foltz, C. B., \& Chaffee, F. H. 1995, AJ 109, 1498

\medskip
\noindent
Hewett, P. C., Foltz, C. B., Chaffee, F. H., Francis, P. J., Weymann, R. J.,
Morris, S. L., Anderson, S. F., \& MacAlpine, G. M. 1991, AJ, 101, 1121

\medskip
\noindent
Hooper, E. J., Impey, C. D., Foltz, C. B., \& Hewett, P. C. 1995, ApJ, 445, 62

\medskip
\noindent
Morris, S. L., Weymann, R. J., Anderson, S. F., Hewett, P. C., Foltz, C. B.,
Chaffee, F. H., Francis, P. J., \& MacAlpine, G. M. 1991, AJ, 102, 1627

\medskip
\noindent
Neugebauer, G., Green, R. F., Matthews, K., Schmidt, M., Soifer, B. T.,
\& Bennet, J. 1987, ApJS, 63, 615

\medskip
\noindent
O'Brien, P. T., Gondhalekar, P. M., \& Wilson, R. 1988, MNRAS, 233, 801

\medskip
\noindent
Richstone, D. O., \& Schmidt, M. 1980, ApJ, 235, 361

\medskip
\noindent
Rocca-Volmerange, B., \& Guiderdoni, B. 1988, A \& AS, 75, 93

\medskip
\noindent
Sanders, D. B., Phinner, E.S., Neugebauer, G., Soifer, B. T., \& Matthews, K.
1989, ApJ, 347, 29

\medskip
\noindent
Sargent, W. L. W., Steidel, C. C., \& Boksenberg, A. 1989, ApJS, 69,
703

\medskip
\noindent
Serjeant, S., \& Rawlings, S. 1996, Nature, 379, 304

\medskip
\noindent
Visnovsky, K. L., Impey, C. D., Foltz, C. B., Hewett, P. C., Weymann, R. J.,
\& Morris, S. L. 1992, ApJ, 391, 560

\medskip
\noindent
Webster, R. L., Francis, P. J., Peterson, B. A., Drinkwater, M. J., \&
Masci, F. J. 1995, Nature, 375, 469

\medskip
\noindent
Wills, B., Netzer, H., \& Wills, D. 1985, ApJ, 288, 94

\newpage

\begin{small}

\begin{center}
{\bf Table 1. Observed Magnitudes}
\end{center}
\begin{tabular}{lccccccl} \\ \hline
Name & $B$ & $V$ & Old & $H$ & $K_s$ & Redshift & Radio  \\
   &     &     &  $B_J$    &     &     &        & Loud?  \\ \hline 
0004$+$0147 & 18.277 & 17.672 & 18.13 & 15.99  & 15.56  & 1.7110 & ?   \\
0006$+$0200 & 17.968 & 17.605 & 16.51 & 16.35  & 15.62  & 2.3358 & No  \\
0006$+$0230 & 18.569 & 18.339 & 17.99 & 16.83  & 16.00  & 2.0981 & No  \\
0007$+$0142 & 18.344 & 18.154 & 18.26 & 16.61  & 16.08  & 1.7654  & Yes \\
0009$+$0219 & 18.360 & 17.960 & 17.99 & 15.91  & \ldots & 2.6416 & No  \\
0010$-$0012 & 19.036 & 18.713 & 18.46 & \ldots & \ldots & 2.1536 & ?   \\
0013$-$0029 & 18.436 & 18.048 & 18.18 & \ldots & \ldots & 2.0835 & No  \\
0018$-$0220 & 17.486 & 17.105 & 17.44 & \ldots & 15.46  & 2.5960 & No  \\ 
0049$+$0045 & 17.766 & 17.540 & 17.50 & 15.74  & \ldots & 2.2647 & No  \\
0052$-$0058 & 18.347 & 18.194 & 17.93 & 16.29  & 15.35  & 2.2120 & No  \\
0056$-$0009 & 17.929 & 17.664 & 17.73 & 15.90  & \ldots & 0.7175 & Yes \\
0106$+$0119 & 18.498 & 18.171 & 18.34 & 15.12  & 14.04  & 2.0989 & Yes \\
0256$-$0000 & 18.876 & 17.606 & 18.22 & 15.72  & 15.19  & 3.3638 & Yes \\
0256$-$0031 & \ldots & \ldots & 17.59 & 15.49  & 14.80  & 1.9951 & ?   \\
0257$-$0254 & \ldots & \ldots & 17.60 & 14.71  & 14.51  & 1.0700 & ?   \\
0258$+$0210 & 18.120 & 17.907 & 17.97 & 15.88  & 15.52  & 2.5239 & No  \\
0301$+$0015 & \ldots & \ldots & 18.31 & 16.44  & \ldots & 1.6458 & ?   \\
0302$-$0019 & 18.409 & 17.436 & 17.78 & 15.87  & 15.27  & 3.2814 & No  \\
1012$+$0213 & \ldots & \ldots & 17.63 & \ldots & 15.10  & 1.3778 & Yes \\
1013$+$0124 & \ldots & \ldots & 16.52 & \ldots & 14.48  & 0.7790 & Yes \\
1215$+$1121 & \ldots & \ldots & 16.61 & \ldots & 14.96  & 1.3984 & Yes \\
1216$+$1754 & \ldots & \ldots & 18.07 & \ldots & 15.23  & 1.8097 & Yes \\
1222$+$1433 & \ldots & \ldots & 17.11 & \ldots & 14.85  & 1.3368 & No  \\
1233$+$1524 & \ldots & \ldots & 18.06 & \ldots & 16.20  & 1.5441 & No  \\
2230$+$0232 & 18.387 & 18.038 & 18.05 & 16.43  & 15.72  & 2.1468 & No  \\
2231$-$0015 & 18.096 & 17.394 & 17.53 & 15.46  & 15.02  & 3.0150 & No  \\
2231$-$0048 & 17.623 & 17.241 & 17.57 & \ldots & \ldots & 1.2095 & ?   \\
2231$-$0212 & 18.357 & 18.065 & 18.15 & \ldots & 15.80  & 1.9052 & ?   \\
2239$+$0007 & 18.796 & 18.453 & 18.28 & 16.32  & \ldots & 1.4401 & ?   \\
2241$+$0014 & \ldots & \ldots & 17.57 & 17.72  & 15.53  & 2.1310 & No  \\
2241$+$0016 & 18.361 & 17.996 & 18.30 & 16.52  & 15.74  & 1.3936 & ?   \\
2243$+$0141 & 18.911 & 18.544 & 18.25 & 16.52  & \ldots & 2.3140 & No  \\
2244$-$0105 & 18.091 & 17.680 & 17.95 & 15.91  & 15.29  & 2.0300 & No  \\
2244$-$0208 & 19.421 & 19.016 & 18.4  & 17.55  & \ldots & 1.9682 & ?   \\
2247$+$0135 & 17.945 & 17.378 & 17.77 & \ldots & \ldots & 1.1284 & ?   \\
2248$+$0127 & 18.582 & 18.163 & 18.21 & \ldots & 15.81  & 2.5585 & No  \\
2350$-$0045B & \ldots & \ldots & 18.38 & \ldots & 15.69 & 0.4437 & ?   \\ 
\hline
\end{tabular}

\end{small}

\newpage

\begin{center}
{\bf Table 2. Magnitude Correction Factors}
\end{center}

\begin{tabular}{lcccc} \\ \hline
Name  & $B$ &  $V$ &  $H$ &  $K$ \\
      & \multicolumn{4}{c}{Correction to Observed Magnitude} \\ \hline 
0004$+$0147 &  0.06 &  0.03 & 0.0    &  0.0     \\
0006$+$0200 &  0.10 &  0.10 & 0.08 & 0.28  \\
0006$+$0230 &  0.13 &  0.05 & 0.08 & 0.27  \\
0007$+$0142 &  0.14 &  0.03 & 0.0    &  0.0     \\
0009$+$0219 &  0.08 &  0.11 & 0.0    & \ldots   \\
0018$-$0220 &  0.12 &  0.07 & \ldots  &  0.0     \\
0049$+$0045 &  0.07 &  0.09 & 0.08 & \ldots   \\
0052$-$0058 &  0.30 &  0.0  & 0.08 & 0.27  \\
0056$-$0009 &  0.08 &  0.0  & 0.0  & \ldots   \\
0106$+$0119 &  0.19 &  0.0  & 0.08 & 0.27  \\
0256$-$0000 & \ldots & 0.57 & 0.0   & 0.08  \\
0258$+$0210 &  0.04 &  0.13 & 0.09 &  0.0     \\
0302$-$0019 & \ldots & 0.48 & 0.0   & 0.08  \\
2230$+$0232 &  0.13 &  0.0  & 0.08 & 0.27  \\
2231$-$0015 & \ldots & 0.0  & 0.0   & 0.08  \\
2231$-$0212 &  0.06 &  0.04 & \ldots  & 0.08  \\
2239$+$0007 &  0.06 &  0.0  & 0.28 & \ldots   \\
2241$+$0016 &  0.06 &  0.0  & 0.28 &   0.0    \\
2243$+$0141 &  0.18 &  0.10 & 0.08 & \ldots   \\
2244$-$0105 &  0.04 &  0.06 & 0.0  & 0.26  \\
2244$-$0208 &  0.09 &  0.11 & 0.0 & \ldots   \\
2248$+$0127 &  0.20 &  0.12 & \ldots  &   0.0    \\ \hline
\end{tabular}

\newpage

\begin{center}
{\bf Table 3. Corrected continuum fluxes}
\end{center}

\begin{tabular}{lcccccc} \\ \hline
Name  & $0.44 \mu$m &  $0.55 \mu$m &  $1.6 \mu$m &  $2.1 \mu$m 
& Residuals (rms) & Slope \\
      & \multicolumn{5}{c}{(flux $\times 10^{-29} {\rm W\ m}^{-2}{\rm Hz}^{-1}$
[mJy])} & $\alpha$ \\ \hline 
0004$+$0147 &  0.205 &  0.294 &  0.438 &  0.411 & 0.035 & $-0.377$ \\
0006$+$0200 &  0.263 &  0.292 &  0.292 &  0.301 & 0.009 & $-0.057$ \\
0006$+$0230 &  0.147 &  0.156 &  0.188 &  0.214 & 0.005 & $-0.220$ \\
0007$+$0142 &  0.180 &  0.189 &  0.248 &  0.255 & 0.035 & $-0.230$ \\
0009$+$0219 &  0.155 &  0.209 &  0.472 & \ldots & 0.004 & $-0.735$ \\
0018$-$0220 &  0.382 &  0.476 & \ldots &  0.451 & 0.041 & $-0.005$ \\
0049$+$0045 &  0.325 &  0.314 &  0.513 & \ldots & 0.016 & $-0.390$ \\
0052$-$0058 &  0.154 &  0.187 &  0.309 &  0.389 & 0.011 & $-0.556$ \\
0056$-$0009 &  0.278 &  0.305 &  0.476 & \ldots & 0.000 & $-0.417$ \\
0106$+$0119 &  0.165 &  0.191 &  0.907 &  1.298 & 0.012 & $-1.368$ \\
0256$-$0000 & \ldots &  0.189 &  0.562 &  0.768 & 0.007 & $-1.048$ \\
0258$+$0210 &  0.223 &  0.215 &  0.446 &  0.426 & 0.030 & $-0.459$ \\
0302$-$0019 & \ldots &  0.240 &  0.489 &  0.499 & 0.027 & $-0.564$ \\
2230$+$0232 &  0.174 &  0.215 &  0.272 &  0.277 & 0.012 & $-0.263$ \\
2231$-$0015 & \ldots &  0.391 &  0.714 &  0.628 & 0.066 & $-0.400$ \\
2231$-$0212 &  0.190 &  0.204 & \ldots &  0.329 & 0.001 & $-0.354$ \\
2239$+$0007 &  0.127 &  0.148 &  0.250 & \ldots & 0.002 & $-0.513$ \\
2241$+$0016 &  0.190 &  0.223 &  0.208 &  0.348 & 0.044 & $-0.286$ \\
2243$+$0141 &  0.103 &  0.124 &  0.250 & \ldots & 0.002 & $-0.675$ \\
2244$-$0105 &  0.249 &  0.284 &  0.472 &  0.415 & 0.036 & $-0.371$ \\
2244$-$0208 &  0.070 &  0.079 &  0.104 & \ldots & 0.002 & $-0.293$ \\
2248$+$0127 &  0.126 &  0.173 & \ldots &  0.326 & 0.007 & $-0.521$ \\ \hline
\end{tabular}

\newpage

\noindent
{\bf Figure Captions:}

\medskip

\noindent
{\bf Figure 1}---The distribution of magnitude changes of QSOs, between
the old, UK Schmidt $B_J$ magnitudes and those presented in this
paper.

\medskip

\noindent
{\bf Figure 2}---The distribution of continuum slopes, as measured by
the power-law index $\alpha$, where the continuum flux per unit wavelength
$f(\nu)$ if fit by a function of the form $f(\nu) \propto \nu^{\alpha}$.

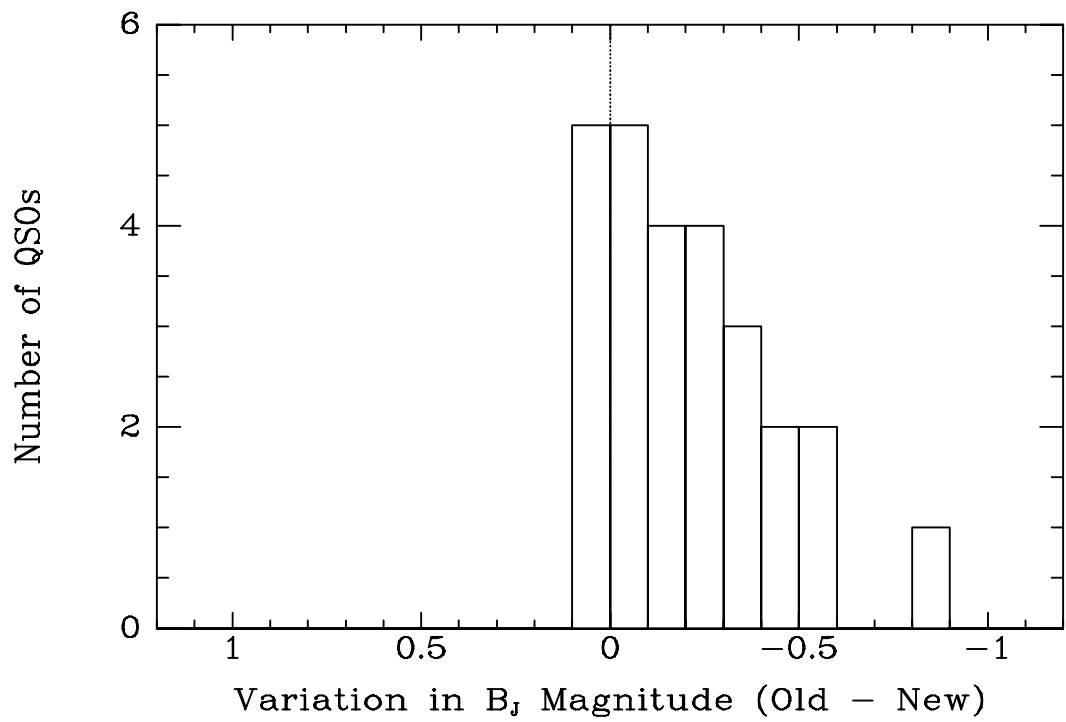
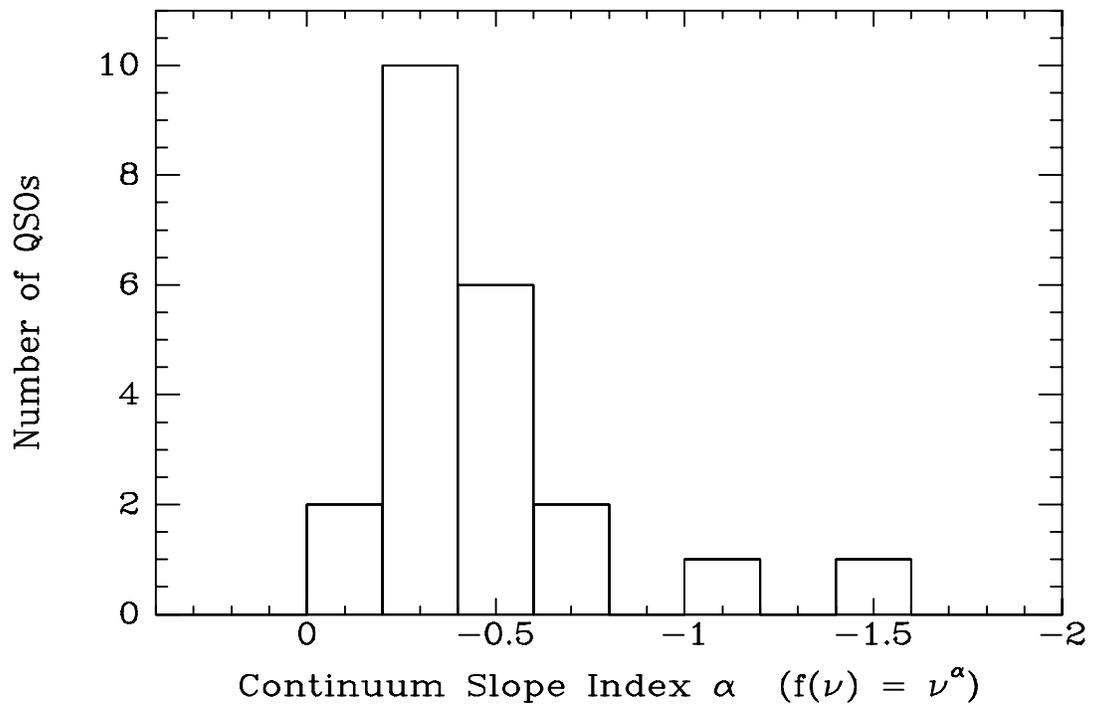
\end{document}